# Nanoscale

## PAPER

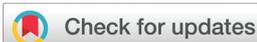

# Differential proteomics highlights macrophage-specific responses to amorphous silica nanoparticles†

Bastien Dalzon,[a] Catherine Aude-Garcia,[a] Véronique Collin-Faure,[a] Hélène Diemer,[b] David Béal,[c] Fanny Dussert,[c] Daphna Fenel,[d] Guy Schoehn,[d] Sarah Cianférani,[b] Marie Carrière[b] and Thierry Rabilloud  *[a]

The technological and economic benefits of engineered nanomaterials may be offset by their adverse effects on living organisms. One of the highly produced nanomaterials under such scrutiny is amorphous silica nanoparticles, which are known to have an appreciable, although reversible, inflammatory potential. This is due to their selective toxicity toward macrophages, and it is thus important to study the cellular responses of this cell type to silica nanoparticles to better understand the direct or indirect adverse effects of nanosilica. We have here studied the responses of the RAW264.7 murine macrophage cells and of the control MPC11 plasma cells to subtoxic concentrations of nanosilica, using a combination of proteomic and targeted approaches. This allowed us to document alterations in the cellular cytoskeleton, in the phagocytic capacity of the cells as well as their ability to respond to bacterial stimuli. More surprisingly, silica nanoparticles also induce a greater sensitivity of macrophages to DNA alkylating agents, such as styrene oxide, even at doses which do not induce any appreciable cell death.

## 1. Introduction

Silica-based particulate materials are highly used as abrasives both in the industry and in consumer products such as toothpastes, as well as reinforcing agents (*e.g.* as mineral charges in tires) or in the high-tech industry (*e.g.* in photovoltaics or as high precision molding agents). These wide uses increase the exposure potential of individuals to silica, which poses in turn the problem of the direct and indirect toxicity of silica. In this frame, crystalline silica has long been known as the causative agent of silicosis and is therefore highly regulated. Conversely, amorphous silica has been demonstrated to induce only a transient and reversible inflammation upon pulmonary exposure,[1–3] and is therefore considered as safer. However, the

[a]Laboratory of Chemistry and Biology of Metals, UMR 5249, Univ. Grenoble Alpes, CNRS, CEA, Grenoble, France. E-mail: thierry.rabilloud@cea.fr
[b]Laboratoire de Spectrométrie de Masse BioOrganique, Université de Strasbourg, CNRS, IPHC UMR 7178, 67000 Strasbourg, France
[c]Chimie Interface Biologie pour l'Environnement, la Santé et la Toxicologie (CIBEST), UMR 5819, Univ. Grenoble Alpes, CEA, CNRS, INAC, SyMMES, F-38000 Grenoble, France
[d]Institut de Biologie Structurale Jean-Pierre Ebel, UMR5075, Univ. Grenoble Alpes, CEA, CNRS, Grenoble, France
†Electronic supplementary information (ESI) available. See DOI: 10.1039/c7nr02140b

amorphous silica-induced inflammation can be pronounced[3] and are found *in vivo* and *in vitro* as well.[4]

Considerable work has been devoted to the analysis of the toxicity of amorphous silica *in vitro*, and strong directions have emerged. One of them is the differential sensitivity of different cell types to amorphous silica,[5–8] and the other one is the influence of the size of the nanoparticles.[5,9–15] However, it has been shown that within a given cell type, the type of response is similar for particles of different sizes.[16]

Regarding the toxic mechanisms induced by amorphous silica, oxidative stress effects have been demonstrated,[17–20] and seems to be linked more to direct ROS generation than to an indirect mechanism *via* glutathione depletion.[21,22] Genotoxicity has also been observed,[23–27] and is likely to be linked to the oxidative stress mentioned above.

Beyond these toxic mechanisms, it is interesting to understand more widely the cellular responses at sub-toxic concentrations of amorphous silica, as they may be linked to the inflammatory responses observed *in vitro*,[8,11,14,17,28–30] or *in vivo*.[2]

One of the emerging mechanisms at play is the autophagy/inflammasome axis,[31–36] which clearly plays a pivotal role in the induction of pro-inflammatory cytokines by silica.

However there are clearly other cellular responses to amorphous silica, as exemplified by transcriptomic studies,[16,37,38] and deciphering these responses may help to understand

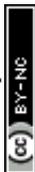

cross toxicities between amorphous silica and metals[39,40] or organic compounds.[41]

In this frame, proteomic studies can be useful in addition to transcriptomics, and have indeed been used to study cellular responses to amorphous silica, on keratinocytes[42] and in lung epithelial cells.[43] We have thus decided to perform a proteomic study of the effects of amorphous silica on macrophages, using the well documented RAW264.7 line.[44] As the cellular responses have been reported to be largely conserved across the size range for silica[16] we have focused our study on a single precipitated silica nanoparticle, previously used in ecotoxicology.[45] In order to take into account the cell type-specific responses, we have performed this study simultaneously on the RAW264.7 macrophage cell line and in a control, less silica-sensitive cell line of the same genotype and of hematopoietic origin, the MPC11 plasmacytoma cell line. This cell line grows at the same speed as the RAW264.7 line in the same culture medium, and also has a similar nucleocytoplasmic ratio.

## 2. Experimental

Most experiments have been performed essentially as described in previous publications.[46–48] Details are given here for the sake of the consistency of the paper. All biological experiments were carried out at least on three independent biological replicates.

### 2.1. Nanoparticles

The silica nanoparticles (Ludox TMA®) were purchased from Sigma, directly as a concentrated suspension. This suspension was diluted to a silica concentration of 1 mg ml$^{-1}$ in distilled water just prior to use. The actual size of the particles was determined after dilution in water or in complete culture medium by dynamic light scattering, using a Wyatt Dynapro Nanostar instrument. The morphology of samples was observed by TEM (Transmission Electron Microscopy). Samples were absorbed to the clean side of a carbon film on mica and transferred to a 400-mesh copper grid. The images were taken under low dose conditions (<10 e$^-$ Å$^{-2}$) at a magnification of 11k×, 13k×, 23k× and 30k× with defocus values between 1.2 and 2.5 μm on a Tecnai 12 LaB6 electron microscope at 120 kV accelerating voltage using a CCD Camera Gatan Orius 1000.

### 2.2. Cell culture

The mouse macrophage cell line RAW264.7 and the mouse plasmacytoma cell line MPC11 were obtained from the European Cell Culture Collection (Salisbury, UK). The cells were cultured in RPMI 1640 medium supplemented with 10% fetal bovine serum (FBS). Cells were seeded at 200 000 cells per ml and harvested at 1 000 000 cells per ml. For treatment with nanoparticles, cells were seeded at 500 000 cells per ml. They were treated with nanoparticles on the following day and harvested after a further 24 hours in culture. Cell viability was measured by a dye exclusion assay, either with eosin

(1 mg ml$^{-1}$)[49] under a microscope or with propidium iodide (1 μg ml$^{-1}$)[50] in a flow cytometry mode. For cross toxicity experiments, the cells were first exposed to silica alone for 6 hours. The tested inhibitors or toxicants were then added for an additional 18 hours and the cell viability measured afterwards.

### 2.3. Phagocytosis and particle internalization assay

The phagocytic activity was measured using fluorescent latex beads (1 μm diameter, green labelled, catalog number L4655 from Sigma). The beads were pre-incubated at a final concentration of 55 μg mL$^{-1}$ for 30 minutes at 37 °C in PBS/FBS (v/v). Then, they were incubated with cells (5 μg mL$^{-1}$) for 2 h 30 min at 37 °C. The cells were harvested and washed with PBS. The cells were resuspended by vortexing with addition of 3/4 water volume and then 1/4 NaCl (35 mg mL$^{-1}$) volume was added under vortexing in order to clean the cell surface of adsorbed particles. The cells were harvested in PBS with propidium iodide (1 μg mL$^{-1}$). Viability and phagocytic activity were measured simultaneously by flow cytometry on a FacsCalibur instrument (Beckton Dickinson). The dead cells (propidium positive) were excluded from the analysis.

For the internalization assay, latex nanoparticles (fluorescent green, from Sigma) were used. The nanoparticles were added directly to the serum-containing cell culture medium and left for 24 hours in contact with the cells. Post-exposure cell harvesting, treatment and analysis were performed similar to the phagocytosis assay.

### 2.4. Mitochondrial transmembrane potential measurement

The mitochondrial transmembrane potential was assessed by Rhodamine 123 uptake. The cells were incubated with Rhodamine 123 (80 nM) for 30 minutes at 37 °C, 5% CO$_2$ then rinsed twice in cold glucose (1 mg mL$^{-1}$)–PBS (PBSG) and harvested in cold PBSG supplemented with propidium iodide (1 μg mL$^{-1}$). The mitochondrial potential of cells was analysed by flow cytometry on a FacsCalibur instrument (Beckton Dickinson). The dead cells (propidium positive) were excluded from the analysis. The low Rhodamine concentration was used to avoid intramitochondrial fluorescence quenching that would result in a poor estimation of the mitochondrial potential.[51]

### 2.5. Enzyme assays

The enzymes were assayed according to published procedures. Isocitrate dehydrogenase was assayed by a coupled assay using nitro blue tetrazolium as the final acceptor and phenazine methosulfate as a relay.[52] Biliverdin reductase was assayed directly for the NADPH-dependent conversion of biliverdin into bilirubin, followed at 450 nm.[53] Lactoylglutathione lyase activity was followed at 240 nm as previously described.[54]

The cell extracts for enzyme assays were prepared by lysing the cells for 20 minutes at 0 °C in 20 mM Hepes (pH 7.5), 2 mM MgCl$_2$, 50 mM KCl, 1 mM EGTA, 0.15% (w/v) tetradecyldimethylammonio propane sulfonate (SB 3–14), followed by centrifugation at 15 000$g$ for 15 minutes to clear the extract. The protein concentration was determined by a dye-binding assay.[55]

## 2.6. NO production and cytokine production

The cells were grown to confluence in a 6 well plate and pre-treated with silica for 6 hours. Then half of the wells were treated with 100 ng ml$^{-1}$ LPS (from salmonella, purchased from Sigma), and arginine monohydrochloride was added to all the wells (5 mM final concentration) to give a high concentration of substrate for the nitric oxide synthase. After 18 hours of incubation, the cell culture medium was recovered, centrifuged at 10 000$g$ for 10 minutes to remove cells and debris, and the nitrite concentration in the supernatants was read at 540 nm after addition of an equal volume of Griess reagent and incubation at room temperature for 30 minutes.

For cytokine production, a commercial kit (BD Cytometric Bead Array, catalog number 552364 from BD Biosciences) was used. The supernatant of cells treated with NP-SiO$_2$ was recovered and the kit protocol was followed.

## 2.7. F-actin staining

The experiments were performed essentially as previously described.[56] The cells were cultured on coverslips placed in 6-well plates and exposed to silica or latex nanoparticles for 24 h at 37 °C. At the end of the exposure time, the cells were washed twice for 5 min at 4 °C in PBS and fixed in 4% paraformaldehyde for 30 min at room temperature. After two washes (5 min/4 °C in PBS), they were permeabilized in 0.1% Triton X-100 for 5 min at room temperature. After two more washes in PBS, 500 nM Phalloidin-Atto 550 (Sigma) was added to the cells and left undisturbed for 20 min at room temperature in the dark. Coverslip-attached cells were washed, placed on microscope slides (Thermo Scientific) using a Vectashield mounting medium containing DAPI (Eurobio) and imaged using a Zeiss LSM 800 confocal microscope. The images were processed using ImageJ software.

## 2.8. RT-qPCR

RNA was extracted using the GenElute™ mammalian total RNA miniprep kit with the optional DNase treatment step, then reverse-transcribed using SuperScript III Reverse Transcriptase (Life Technologies). RNA concentration and purity were assessed by measuring Abs 260/Abs 280 and Abs 260/Abs 230 absorbance ratios using a Nanodrop ND-1000 spectrophotometer (Thermo Fisher Scientific). Then, cDNA from each of the three biological replicates for each exposure condition was loaded in duplicate on a 96-well plate. Primer sequences are given in ESI Table 2.† Their efficiencies were experimentally checked for compliance using a mix of all samples, with a quality criterion of 2 ± 0.3. Quantitative PCR was performed on a MX3005P Multiplex Quantitative PCR thermocycler (Stratagene), using the following thermal cycling steps: 95 °C for 5 min, then 95 °C for 15 s, 55 °C for 20 s and 72 °C for 40 s 40 times and finally 95 °C for 1 min, 55 °C for 30 s and 95 °C for 30 s for the dissociation curve. $C_q$ was determined using the Mx-Pro 3.20 software with default settings. Glyceraldehyde-3-phosphate dehydrogenase (GAPDH) and 18S ribosomal 1 (S18) were chosen as reference genes for normalization, validated using BestKeeper.[57] mRNA expression analysis, normalization and statistical analysis were performed using REST2009 software[58] using the $\Delta\Delta C_q$ method and a pair-wise fixed reallocation randomization test.

## 2.9. Proteomics

The 2D gel based proteomic experiments were essentially carried out as previously described,[46] at least on independent biological triplicates. However, detailed materials and methods are provided for the sake of paper consistency.

**2.9.1. Sample preparation.** The cells were collected by scraping, and then washed three times in PBS. The cells were then washed once in TSE buffer (10 mM Tris-HCl (pH 7.5), 0.25 M sucrose, 1 mM EDTA), and the volume of the cell pellet was estimated. The pellet was resuspended in its own volume of TSE buffer. Then 4 volumes (respective to the cell suspension just prepared) of concentrated lysis buffer (8.75 M urea, 2.5 M thiourea, 5% w/v CHAPS, 6.25 mM TCEP-HCl, 12.5 mM spermine base) were added and the solution was left undisturbed for extraction at room temperature for 1 hour. The nucleic acids were then pelleted by ultracentrifugation (270 000$g$ at room temperature for 1 h), and the protein concentration in the supernatant was determined by a dye-binding assay.[55] Carrier ampholytes (Pharmalytes pH 3–10) were added to a final concentration of 0.4% (w/v), and the samples were kept frozen at −20 °C until use.

**2.9.2. Isoelectric focusing.** Home-made 160 mm long 4–8 linear pH gradient gels[59] were cast according to published procedures.[60] Four mm-wide strips were cut, and rehydrated overnight with the sample, diluted in a final volume of 0.6 ml of rehydration solution (7 M urea, 2 M thiourea, 4% CHAPS, 0.4% carrier ampholytes (Pharmalytes pH 3–10) and 100 mM dithiodiethanol).[61]

The strips were then placed in a Multiphor plate (GE Healthcare), and IEF was carried out with the following electrical parameters: 100 V for 1 hour, then 300 V for 3 hours, then 1000 V for 1 hour, then 3400 V up to 60–70 kVh. After IEF, the gels were equilibrated for 20 minutes in 125 mM Tris, 100 mM HCl, 2.5% SDS, 30% glycerol and 6 M urea.[62] They were then transferred on top of the SDS gels and sealed in place with 1% agarose dissolved in 125 mM Tris, 100 mM HCl, 0.4% SDS and 0.005% (w/v) bromophenol blue.

**2.9.3. SDS electrophoresis and protein detection.** Ten percent gels (160 × 200 × 1.5 mm) were used for protein separation. The Tris taurine buffer system[63] was used and operated at an ionic strength of 0.1 and a pH of 7.9. The final gel composition is thus 180 mM Tris, 100 mM HCl, 10% (w/v) acrylamide, and 0.27% bisacrylamide. The upper electrode buffer is 50 mM Tris, 200 mM Taurine, and 0.1% SDS. The lower electrode buffer is 50 mM Tris, 200 mM glycine, and 0.1% SDS. The gels were run at 25 V for 1 hour, then 12.5 W per gel until the dye front has reached the bottom of the gel. Detection was carried out by tetrathionate silver staining.[64]

**2.9.4. Image analysis.** The gels were scanned after silver staining on a flatbed scanner (Epson perfection V750), using a 16 bit grayscale image acquisition. The gel images were then analyzed using the Delta 2D software (v 3.6). Spots that were never expressed above 100 ppm of the total spots were first filtered out. Then, significantly-varying spots were selected on the basis of their Student's $t$-test $p$-value between the treated and the control groups. Spots showing a $p$-value lower than 0.05 were selected. This strategy is used to avoid the use of arbitrary thresholds, which can result in discarding statistically-valid relevant changes and including non-valid changes.[65] The false positive concern arising from the multiple testing problem was addressed using the Storey–Tibshirani approach,[66] as classical statistical filters (e.g. Bonferroni or Benjamini–Hochberg) yield to over-rejection of valid results.[67] Furthermore, we checked that all the spots that we found through the $t$-test also had a $p < 0.05$ in a non-parametric Mann–Whitney $U$-test.

**2.9.5. Mass spectrometry.** The spots selected for identification were excised from silver-stained gels and destained with ferricyanide/thiosulfate on the same day as silver staining in order to improve the efficiency of the identification process.[68,69] Gel digestion was performed with an automated protein digestion system, MassPrep Station (Waters, Milford, USA). The gel plugs were washed twice with 50 µL of 25 mM ammonium hydrogen carbonate ($NH_4HCO_3$) and 50 µL of acetonitrile. The cysteine residues were reduced by 50 µL of 10 mM dithiothreitol at 57 °C and alkylated by 50 µL of 55 mM iodoacetamide. After dehydration with acetonitrile, the proteins were cleaved in gel with 10 µL of 12.5 ng µL$^{-1}$ of modified porcine trypsin (Promega, Madison, WI, USA) in 25 mM $NH_4HCO_3$. The digestion was performed overnight at room temperature. The generated peptides were extracted with 30 µL of 60% acetonitrile in 0.1% formic acid. Acetonitrile was evaporated under vacuum before nanoLC-MS/MS analysis.

NanoLC-MS/MS analysis was performed using a nanoACQUITY Ultra-Performance-LC (Waters Corporation, Milford, USA) coupled to the Synapt™ High Definition Mass Spectrometer™ (Waters Corporation, Milford, USA), or to the TripleTOF 5600 (Sciex, Ontario, Canada).

The nanoLC system was composed of an ACQUITY UPLC® CSH130 C18 column (250 mm × 75 µm with a 1.7 µm particle size, Waters Corporation, Milford, USA) and a Symmetry C18 precolumn (20 mm × 180 µm with a 5 µm particle size, Waters Corporation, Milford, USA). The solvent system consisted of 0.1% formic acid in water (solvent A) and 0.1% formic acid in acetonitrile (solvent B). 4 µL of sample were loaded into the enrichment column for 3 min at 5 µL min$^{-1}$ with 99% of solvent A and 1% of solvent B. Elution of the peptides was performed at a flow rate of 300 nL min$^{-1}$ with a 8–35% linear gradient of solvent B in 9 minutes.

The Synapt™ High Definition Mass Spectrometer™ (Waters Corporation, Milford, USA) was equipped with a Z-spray ion source and a lock mass system. The system was fully controlled using MassLynx 4.1 SCN639 (Waters Corporation, Milford, USA). The capillary voltage was set at 2.8 kV and the cone voltage at 35 V. Mass calibration of the TOF was achieved using fragment ions from Glu-fibrino-peptide B on the [50;2000] $m/z$ range. Online correction of this calibration was performed with Glu-fibrino-peptide B as the lock-mass. The ion $(M + 2H)^{2+}$ at $m/z$ 785.8426 was used to calibrate MS data and the fragment ion $(M + H)^+$ at $m/z$ 684.3469 was used to calibrate MS/MS data during the analysis.

For tandem MS experiments, the system was operated with automatic switching between MS (0.5 s per scan on $m/z$ range [150;1700]) and MS/MS modes (0.5 s per scan on $m/z$ range [50;2000]). The two most abundant peptides (intensity threshold 20 counts per s), preferably doubly and triply charged ions, were selected on each MS spectrum for further isolation and CID fragmentation using collision energy profile. Fragmentation was performed using argon as the collision gas.

Mass data collected during analysis were processed and converted into .pkl files using ProteinLynx Global Server 2.3 (Waters Corporation, Milford, USA). Normal background subtraction type was used for both MS and MS/MS with 5% threshold and polynomial correction of order 5. Smoothing was performed on MS/MS spectra (Savitsky-Golay, 2 iterations, window of 3 channels). Deisotoping was applied for MS (medium deisotoping) and for MS/MS (fast deisotoping).

The TripleTOF 5600 (Sciex, Ontario, Canada) was operated in positive mode, with the following settings: ionspray voltage floating (ISVF) 2300 V, curtain gas (CUR) 10, interface heater temperature (IHT) 150, ion source gas 1 (GS1) 2, declustering potential (DP) 80 V. Information-dependent acquisition (IDA) mode was used with Top 10 MS/MS scans. The MS scan had an accumulation time of 250 ms in $m/z$ [400;1250] range and the MS/MS scans 100 ms in $m/z$ [150;1800] range in high sensitivity mode. Switching criteria were set to ions with a charge state of 2–4 and an abundance threshold of more than 500 counts and exclusion time was set at 4 s. IDA rolling collision energy script was used for automatically adapting the CE. Mass calibration of the analyser was achieved using peptides from digested BSA. The complete system was fully controlled by AnalystTF 1.7 (Sciex). Raw data collected were processed and converted with MSDataConverter in .mgf peak list format.

For protein identification, the MS/MS data were interpreted using a local Mascot server with the MASCOT 2.4.1 algorithm (Matrix Science, London, UK) against UniProtKB/SwissProt (version 2016_01, 550299 sequences). Research was carried out in all species. Spectra were searched with a mass tolerance of 15 ppm for MS and 0.05 Da for MS/MS data, allowing a maximum of one trypsin missed cleavage. Carbamidomethylation of cysteine residues and oxidation of methionine residues were specified as variable modifications. Protein identifications were validated with at least two peptides with a Mascot ion score above 30.

# 3. Results

## 3.1. Nanoparticles characterization and determination of the effective doses

The amorphous silica nanoparticle used (Ludox™ TMA) was characterized by several methods, and the results are summarized in Fig. 1 (panels A–C). Spherical nanoparticles were obtained (primary diameter 26 ± 4 nm), and the average hydrodynamic diameter of these particles, as measured by DLS, was 37 ± 1 nm, with a polydispersity index of 20%. When placed in serum-containing culture medium the hydrodynamic diameter immediately increased to 155 ± 7 nm (polydispersity index 25%). The aggregation state increased over time in the serum-containing medium to up to 365 nm (multimodal) after 24 hours in the medium. Toxicity curves were then determined on the two cell lines of interest. The $LD_{20}$, i.e. the concentration inducing a 20% cell death after 24 hours of treatment, was determined to be 20 μg ml$^{-1}$ for the RAW264 cell line and 100 μg ml$^{-1}$ for the MPC11 cell line (Fig. 1D). These concentrations were chosen for the subsequent studies, as the $LD_{20}$ offers a good compromise between cell viability and biological effect.

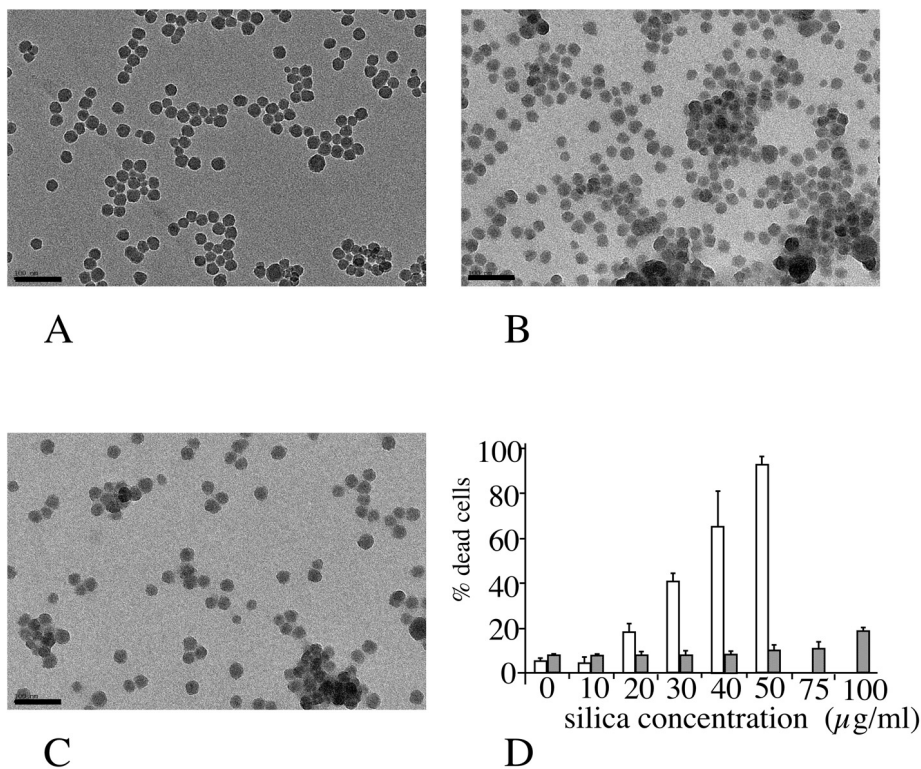

**Fig. 1** Nanoparticles characterization and effects on cell viability. Panel A: TEM image of Ludox TMA silica nanoparticles suspended in water. Panels B and C: TEM images of Ludox TMA silica nanoparticles suspended in complete culture medium (RPMI 1640 + fetal bovine serum). B: Time in complete medium 30 minutes. C: Time in complete medium 24 hours. Scale bar: 100 μm. Panel D: Effects of Ludox TMA on cell viability on the RAW264.7 cell line (white bars) and MPC11 cell line (dotted bars).

In order to determine whether the differences in toxicity were due to different internalization between the two cell lines of interest, we tested the internalization of particles in the two cell lines by using fluorescent latex beads of different diameters (30 and 100 nm). The results, displayed in ESI Fig. 1 and Table 1,† show that silica toxicity parallels the internalization capacity of the cells.

### 3.2. Proteomic studies

In order to gain further insights into the molecular responses of cells to the amorphous silica nanoparticles, we performed proteomic studies. We used two different doses for each cell line. For the RAW264 cell line, 20 μg ml$^{-1}$ (i.e. the $LD_{20}$) and 10 μg ml$^{-1}$ (i.e. a dose where no increased mortality and very minimal functional effects were observed) were used. For the MPC11 cell line we used 20 μg ml$^{-1}$ (i.e. the same dose as on RAW264, but with no visible macroscopic effects in this case) and 100 μg ml$^{-1}$ (i.e. the $LD_{20}$ for this cell line). This proteomic analysis probed 2590 protein species for the RAW264 cell line, and 2180 for the MPC11 cell line. The median coefficient of variation of the spots was 24.5% for the RAW cell line, and 21% for the MPC11 cell line. These coefficients of variation are in the range of those found in typical 2D DIGE experiments, where coefficient of variations range from 18 to 28%, depending on the sample.[70–73] The significant protein changes were detected through the use of a variance-based screen, which compensates automatically for the variability of each spot, and enables to take into account small but reproducible changes, thus avoiding the arbitrary exclusion of changes that can be biologically meaningful. Through this proteomic screen, we could detect modulation of proteins belonging to various functional classes, as shown in Fig. 2, Table 1, ESI Table 2 and Fig. 2–7.† Among the 113 significantly variable spots, 15 were common between RAW264 and MPC11 and 19 varied for the two doses of silica in the RAW264 cell line. This means in turn that the majority of significant variables are specific for the

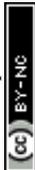

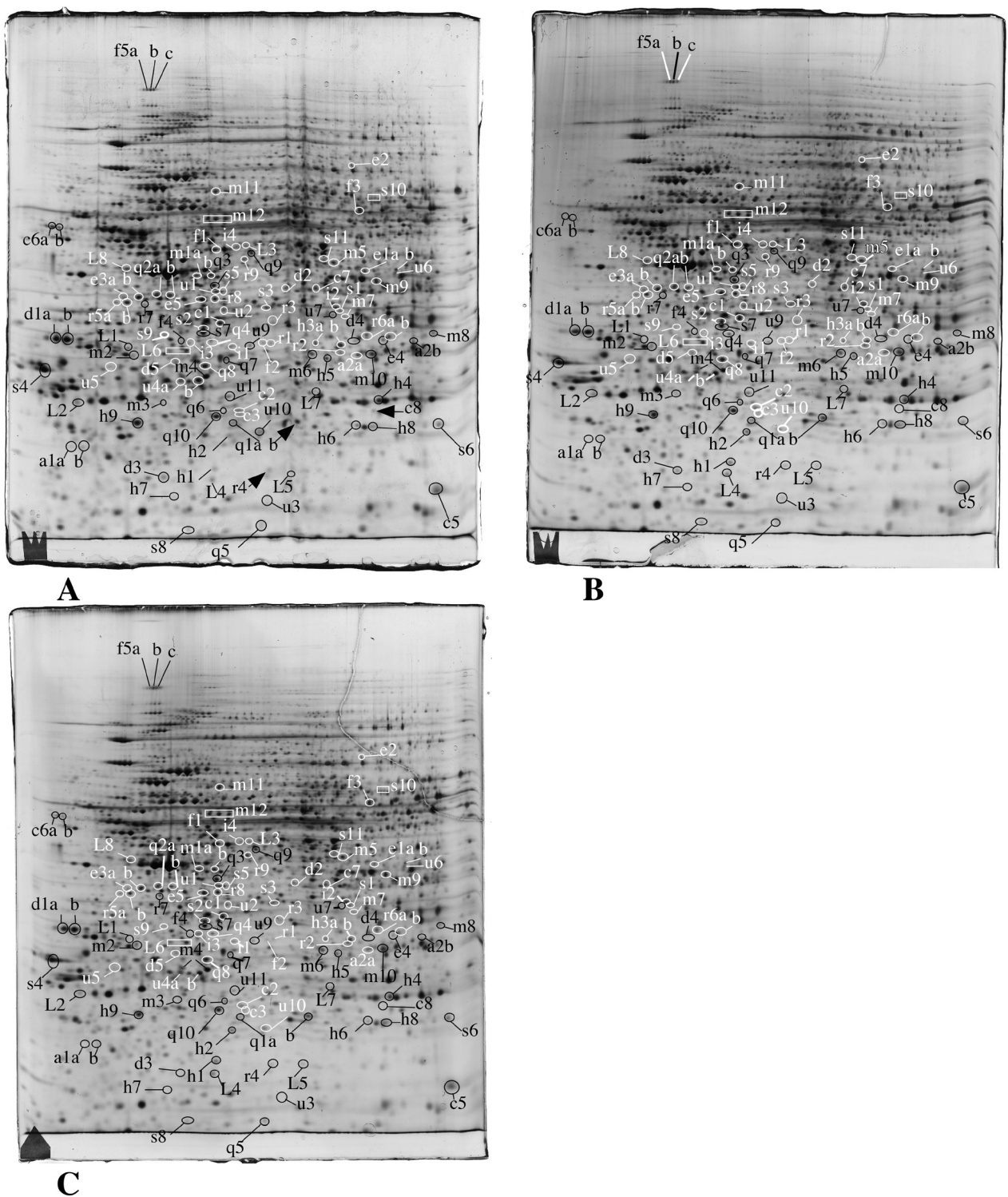

**Fig. 2** Proteomic analysis of total cell extracts by 2D electrophoresis. Total cell extracts of RAW274.7 cells were separated by two-dimensional gel electrophoresis. The first dimensions covered a 4–8 pH range and the second dimension a 15–200 kDa range. Total cellular proteins (150 µg) were loaded on the first dimension gel. A: Gel obtained from control cells. B: Gel obtained from cells treated for 24 hours with 10 µg ml$^{-1}$ Ludox TMA. C: Gel obtained from cells treated for 24 hours with 20 µg ml$^{-1}$ Ludox TMA. The lines and arrows point to spots that show reproducible and statistically significant changes between the control and nanoparticle-treated cells and to the control neighbor spots in some cases. Spot numbering according to Table 1.

**Table 1** Differentially-expressed proteins identified in the proteomic screen

| Spot Id. | Protein Short name | Protein Full name | Prot. acc. no. (uniprot) | Ratio RAW264.7 TMA10/ctl | t test RAW264.7 TMA10 vs. ctl | Ratio RAW264.7 TMA20/ctl | t test RAW264.7 TMA20 vs. ctl | Ratio MPC11 TMA20/ctl | t test MPC11 TMA20 vs. ctl | Ratio MPC11 TMA100/ctl | t test MPC11 vs. ctl |
|---|---|---|---|---|---|---|---|---|---|---|---|
| **Protein quality control & degradation (q)** | | | | | | | | | | | |
| q1a | psb3 ac. | Proteasome beta 3 subunit | Q9R1P1 | 1.01 | 0.94 | 0.65 | 0.01 | 1.09 | 0.77 | 0.93 | 0.87 |
| q1b | psb3 bas | Proteasome beta 3 subunit | Q9R1P1 | 1.31 | 0.01 | 1.29 | 0.04 | 0.91 | 0.41 | 0.71 | 0.07 |
| q2a | sae1 ac | SUMO-activating enzyme subunit 1 | Q9RIT2 | 0.97 | 0.66 | 0.72 | 0.01 | 1.10 | 0.74 | 1.00 | 1.00 |
| q2b | sae1 bas | SUMO-activating enzyme subunit 1 | Q9RIT2 | 0.9 | 0.51 | 0.33 | 0.01 | 1.07 | 0.60 | 0.85 | 0.14 |
| q3 | psd13 | Proteasome regulatory subunit 13 | Q9WVJ2 | 1.05 | 0.55 | 1.32 | 0.02 | 1.32 | 0.02 | 1.31 | 0.08 |
| q4 | brcc3 | Lys63-deubiquitinase brcc36 | P46737 | 0.96 | 0.65 | 0.82 | 0.02 | 1.62 | 0.09 | 1.51 | 0.04 |
| q5 | ube2n | Ubiquitin conjugating enzyme 2N | P61088 | 1.09 | 0.82 | 0.56 | 0.02 | 0.87 | 0.6 | 0.98 | 0.93 |
| **q6** | **psd10** | **Proteasome regulatory subunit 10** | Q9Z2X2 | **1.46** | 0.11 | **1.73** | **0.03** | 0.95 | 0.75 | **1.37** | **0.03** |
| q7 | psmd8 | Proteasome regulatory subunit 8 | Q9CX56 | 0.84 | 0.23 | 0.69 | 0.03 | 1.14 | 0.04 | 1.01 | 0.9 |
| q8 | psme2 | Proteasome activator complex subunit 2 | P97372 | 1.18 | 0.06 | 1.21 | 0.04 | 1.28 | 0.01 | 1.07 | 0.44 |
| q9 | prs7 | Proteasome regulatory subunit 7 | P46471 | 1.07 | 0.52 | 1.16 | 0.04 | 1.12 | 0.31 | 1.2 | 0.23 |
| **q10** | **psb4** | **Proteasome beta 4 subunit** | P99026 | **1.13** | **0.04** | **1.15** | **0.05** | **1.13** | **0.01** | **1.15** | **0.01** |
| **Protein production & folding (f)** | | | | | | | | | | | |
| f1 | if2b | Translation initiation factor 2 subunit 2 | Q99L45 | 0.92 | 0.51 | 1.50 | 0.01 | 1.08 | 0.62 | 0.69 | 0.02 |
| f2 | dnjc9 | Dnaj homolog subfamily C member 9 | Q91WN1 | 0.71 | 0.15 | 0.43 | 0.02 | 1.09 | 0.48 | 1.00 | 0.99 |
| f3 | tcpz | T-complex protein 1 subunit zeta | P80317 | 0.83 | 0.14 | 0.69 | 0.03 | 0.90 | 0.17 | 1.06 | 0.62 |
| f4 | ppie | Peptidyl-prolyl cis-trans isomerase E | Q9QZH3 | 1.2 | 0.31 | 2.03 | 0.04 | 0.92 | 0.56 | 1.32 | 0.08 |
| f5a* | hyou1 ac | Hypoxia up-regulated protein 1 | Q9JKR6 | 1.46 | 0.24 | 1.63 | 0.10 | 0.29 | 0.03 | 0.55 | 0.23 |
| f5b | hyou1 med | Hypoxia up-regulated protein 1 | Q9JKR6 | 1.28 | 0.27 | 1.71 | 0.03 | 0.73 | 0.16 | 1.08 | 0.78 |
| f5c | hyou1 bas | Hypoxia up-regulated protein 1 | Q9JKR6 | 1.00 | 0.98 | 1.76 | 0.04 | 1.12 | 0.31 | 1.59 | 0.06 |
| **Homeostasis (h)** | | | | | | | | | | | |
| h1 | frih | Ferritin heavy chain | P09528 | 1.78 | 0.02 | 1.88 | 0.01 | 1.28 | 0.02 | 1.34 | 0.08 |
| h2* | fril | Ferritin light chain | P29391 | 1.38 | 0.09 | 1.25 | 0.22 | 0.61 | 0.12 | 0.74 | 0.42 |
| h3a | blvra ac | Biliverdin reductase A | Q9CY64 | 0.84 | 0.27 | 0.67 | 0.01 | 1.06 | 0.61 | 1.08 | 0.66 |
| h3b* | blvra bas | Biliverdin reductase A | Q9CY64 | 1.02 | 0.87 | 1.19 | 0.23 | 1.03 | 0.92 | 1.10 | 0.73 |
| h4 | blvrb | Biliverdin reductase B | Q923D2 | 1.05 | 0.49 | 1.30 | 0.01 | N.D. | N.D. | N.D. | N.D. |
| h5 | nmrl1 | NmrA-like family domain-containing protein 1 | Q8K2T1 | 0.77 | 0.14 | 0.80 | 0.02 | 1.07 | 0.84 | 1.30 | 0.41 |
| h6 | pddc1 | Parkinson disease 7 domain-containing protein | Q8BFQ8 | 0.9 | 0.31 | 0.81 | 0.02 | 1.27 | 0.06 | 1.23 | 0.07 |
| h7 | txd12 | Thioredoxin domain-containing protein 12 | Q9CQU0 | 0.9 | 0.62 | 0.57 | 0.04 | 1.08 | 0.68 | 0.96 | 0.85 |
| h8 | prdctox | Peroxiredoxin 1, oxidized form | P35700 | 1.00 | 0.99 | 1.38 | 0.04 | 0.60 | 0.03 | 0.75 | 0.24 |
| h9 | lgul | Lactoylglutathione lyase | Q9CPU0 | 0.94 | 0.36 | 0.81 | 0.05 | 1.25 | 0.01 | 1.18 | 0.11 |
| **Energy & lipid metabolism (e)** | | | | | | | | | | | |
| e1a | idhc ac | Isocitrate dehydrogenase, cytoplasmic | O88844 | 0.95 | 0.48 | 0.75 | 0.01 | 1.08 | 0.55 | 0.87 | 0.30 |
| e1b* | idhc bas | Isocitrate dehydrogenase, cytoplasmic | O88844 | 0.94 | 0.75 | 1.07 | 0.70 | 0.78 | 0.30 | 0.75 | 0.35 |
| e2 | pfkal | 6-Phosphofructokinase, liver type | P12382 | 0.99 | 0.97 | 1.45 | 0.03 | 0.92 | 0.73 | 0.73 | 0.12 |
| e3a | galK ac | Galactose kinase | Q9R0N0 | 0.75 | 0.03 | 0.74 | 0.05 | 0.94 | 0.42 | 0.86 | 0.22 |
| e3b* | galK bas | Galactose kinase | Q9R0N0 | 0.96 | 0.60 | 0.94 | 0.44 | 1.21 | 0.02 | 1.15 | 0.14 |
| e4 | pipnb | Phosphatidylinositol transfer protein beta isoform | P53811 | 0.53 | 0.02 | 0.52 | 0.02 | 1.17 | 0.56 | 0.91 | 0.70 |
| e5 | fpps bas | Farnesyl pyrophosphate synthase | Q920E5 | 1.04 | 0.60 | 1.21 | 0.03 | 1.15 | 0.03 | 0.90 | 0.23 |
| **DNA metabolism and repair (d)** | | | | | | | | | | | |
| d1a | pcna ac | Proliferating cell nuclear antigen | P17918 | 0.89 | 0.06 | 0.81 | 0.01 | 1.11 | 0.23 | 1.14 | 0.12 |
| d1b | pcna bas | Proliferating cell nuclear antigen | P17918 | 0.90 | 0.13 | 0.89 | 0.09 | 1.27 | 0.02 | 1.23 | 0.02 |

**Table 1** (Contd.)

| Spot Id. | Protein Short name | Protein Full name | Prot. acc. no. (uniprot) | Ratio RAW264.7 TMA10/ctl | t test RAW264.7 TMA10 vs. ctl | Ratio RAW264.7 TMA20/ctl | t test RAW264.7 TMA20 vs. ctl | Ratio MPC11 TMA20/ctl | t test MPC11 TMA20 vs. ctl | Ratio MPC11 TMA100/ctl | t test MPC11 TMA100/ TMA10 vs. ctl |
|---|---|---|---|---|---|---|---|---|---|---|---|
| d2 | rfc2 | Replication factor C subunit 2 | Q9WUK4 | 0.51 | 0.02 | 0.54 | 0.03 | 1.16 | 0.50 | 1.57 | 0.03 |
| d3 | dnph1 | 2'-deoxynucleoside 5'-phosphate N-hydrolase 1 | Q80YJ3 | 0.87 | 0.28 | 0.73 | 0.04 | 1.27 | 0.03 | 1.23 | 0.21 |
| d4 | tatd3 | Putative deoxyribonuclease TATDN3 | Q3U1C6 | 0.59 | 0.06 | 0.53 | 0.04 | 1.13 | 0.21 | 1.10 | 0.64 |
| d5 | rfa2 | Replication protein A 32 kDa subunit | Q62193 | 0.91 | 0.20 | 0.80 | 0.01 | 0.74 | 0.02 | 0.98 | 0.85 |
| **RNA and nucleotide metabolism (r)** | | | | | | | | | | | |
| r1 | tsnax | Translin-associated protein X | Q9QZE7 | 0.82 | 0.26 | 0.57 | 0.02 | 1.28 | 0.24 | 0.67 | 0.17 |
| r2 | samp | SAP domain-containing ribonucleoprotein | Q9DJ13 | 0.76 | 0.08 | 0.64 | 0.02 | 1.27 | 0.10 | 1.12 | 0.35 |
| r3 | osgep | Probable tRNA N6-adenosine threonylcarbamoyltransferase | Q8BWU5 | 0.82 | 0.20 | 0.69 | 0.02 | 1.02 | 0.94 | 0.90 | 0.68 |
| r4 | bt3L4 | Transcription factor BTF3 homolog 4 | Q9CQH7 | 0.18 | 0.03 | 0.15 | 0.03 | **1.36** | **0.02** | **1.24** | **0.32** |
| r5a | **strap ac** | **Serine-threonine kinase receptor-associated protein** | **Q9Z1Z2** | **0.72** | **0.08** | **0.64** | **0.03** | **0.84** | **0.37** | **0.39** | **0.03** |
| r5b* | strap bas | Serine-threonine kinase receptor-associated protein | Q9Z1Z2 | 0.90 | 0.40 | 0.86 | 0.28 | 1.30 | 0.03 | 1.12 | 0.25 |
| r6a | prtps1 a | Ribose-phosphate pyrophosphokinase 1 | Q9D7G0 | 1.01 | 0.91 | 1.61 | 0.04 | 1.09 | 0.3 | 1.24 | 0.12 |
| r6b | prtps1 b | Ribose-phosphate pyrophosphokinase 1 | Q9D7G0 | 0.73 | 0.01 | 0.89 | 0.48 | 1.13 | 0.42 | 1.26 | 0.19 |
| r7 | pihd1 | PIH1 domain-containing protein 1 | Q9WTM5 | 0.49 | 0.02 | 0.42 | 0.02 | 0.43 | 0.03 | 1.27 | 0.46 |
| r8 | **bpnt1** | **3'(2'),5'-Bisphosphate nucleotidase 1** | **Q9Z0S1** | **1.12** | **0.31** | **1.40** | **0.01** | **1.29** | **0.01** | **1.22** | **0.09** |
| r9 | adk | Adenosine kinase | P55264 | 0.68 | 0.01 | 0.50 | 0.01 | 0.82 | 0.21 | 0.97 | 0.80 |
| **Cytoskeleton (c)** | | | | | | | | | | | |
| c1 | caza2 | F-actin-capping protein subunit alpha-2 | P47754 | 0.97 | 0.81 | 0.38 | 0.01 | 0.74 | 0.12 | 1.02 | 0.92 |
| c2 | rhoA | Transforming protein RhoA | Q9QUI0 | 0.80 | 0.07 | 0.47 | 0.01 | 0.91 | 0.45 | 0.95 | 0.74 |
| c3 | rab14 | Ras-related protein Rab-14 | Q91 V41 | 0.80 | 0.07 | 0.47 | 0.01 | 1.01 | 0.91 | 0.97 | 0.75 |
| c4 | twf1 | Twinfilin-1 | Q91YR1 | 1.99 | 0.02 | 2.05 | 0.01 | 1.15 | 0.50 | 0.70 | 0.11 |
| c5 | cof1 1PO4 | Cofilin-1, monophosphorylated form | P18760 | 0.90 | 0.39 | 0.53 | 0.01 | 0.66 | 0.09 | 0.62 | 0.11 |
| c6a | lsp1 ac | Lymphocyte-specific protein 1 | P19973 | 0.68 | 0.02 | 0.82 | 0.11 | N.D. | N.D. | N.D. | N.D. |
| c6b | lsp1 bas | Lymphocyte-specific protein 1 | P19973 | 0.63 | 0.05 | 0.56 | 0.04 | 0.51 | 0.05 | 2.03 | 0.07 |
| c7 | twf2 ac | Twinfilin-2 | Q920P5 | 0.93 | 0.39 | 0.74 | 0.04 | 0.71 | 0.06 | 1.13 | 0.66 |
| c8 | rras | Ras-related protein R-Ras | P10833 | 0.57 | 0.40 | 0.23 | 0.01 | 0.80 | 0.23 | 1.31 | 0.21 |
| **Mitochondria (m)** | | | | | | | | | | | |
| m1a | sucb2 ac | Succinate-CoA ligase [GDP-forming] subunit beta, mitochondrial | Q9Z2I8 | 0.99 | 0.93 | 1.35 | 0.01 | 0.86 | 0.09 | 0.79 | 0.07 |
| m1b | sucb2 bas | Succinate-CoA ligase [GDP-forming] subunit beta, mitochondrial | Q9Z2I8 | 1.17 | 0.01 | 1.19 | 0.01 | 1.19 | 0.16 | 1.13 | 0.45 |
| m2 | **coq9** | **Ubiquinone biosynthesis protein COQ9, mitochondrial** | **Q8K1Z0** | **1.14** | **0.28** | **1.61** | **0.01** | **1.69** | **0.02** | **1.49** | **0.05** |
| m3 | mtx2 | Metaxin-2 | O88441 | 0.68 | 0.01 | 0.48 | 0.01 | 0.86 | 0.46 | 1.18 | 0.10 |
| m4a | plb ac. | Prohibitin | P67778 | 1.02 | 0.72 | 1.30 | 0.03 | 0.73 | 0.15 | 0.75 | 0.31 |
| m4b | **phb** | **Prohibitin** | **P67778** | **1.11** | **0.26** | **1.32** | **0.02** | **1.12** | **0.28** | **1.26** | **0.03** |
| m5 | eftu | Elongation factor Tu, mitochondrial | Q8BFR5 | 1.03 | 0.71 | 1.16 | 0.02 | 1.30 | 0.08 | 1.42 | 0.08 |
| m6 | thtm | 3-Mercaptopyruvate sulfurtransferase | Q99JT9 | 1.14 | 0.05 | 1.24 | 0.02 | 0.92 | 0.44 | 0.90 | 0.39 |
| m7 | htra2 | Serine protease HTRA2, mitochondrial | Q9JIY5 | 0.74 | 0.34 | 1.74 | 0.04 | 1.27 | 0.09 | 1.08 | 0.67 |
| m8 | **hmgCL** | **Hydroxymethylglutaryl-CoA lyase, mitochondrial** | **P38060** | **1.00** | **0.99** | **0.66** | **0.04** | **1.05** | **0.58** | **0.91** | **0.04** |
| m9 | acadl | Long-chain specific acyl-CoA dehydrogenase, mitochondrial | P51174 | 1.00 | 0.97 | 1.15 | 0.04 | 1.13 | 0.25 | 1.13 | 0.2 |
| m10 | vdac2 | Voltage-dependent anion-selective channel protein 2 | Q60930 | 0.95 | 0.59 | 1.16 | 0.05 | 0.86 | 0.19 | 0.91 | 0.28 |

Table 1 (Contd.)

| Spot Id. | Protein Short name | Protein Full name | Prot. acc. no. (uniprot) | Ratio RAW264.7 TMA10/ctl | t test RAW264.7 TMA10 vs. ctl | Ratio RAW264.7 TMA20/ctl | t test RAW264.7 TMA20 vs. ctl | Ratio MPC11 TMA20/ctl | t test MPC11 TMA20 vs. ctl | Ratio MPC11 TMA100/ctl | t test MPC11 TMA100 vs. ctl |
|---|---|---|---|---|---|---|---|---|---|---|---|
| m11 | odp2 ac | Dihydrolipoyllysine-residue acetyltransferase component of pyruvate dehydrogenase complex, mitochondrial | Q8BMF4 | 1.10 | 0.39 | 1.29 | 0.05 | 1.11 | 0.09 | 0.97 | 0.77 |
| **Lysosomes &vesicles (l.)** | | | | | | | | | | | |
| L1 | catZ | Cathepsin Z | Q9WUU7 | 0.25 | 0.01 | 0.10 | 0.01 | 1.49 | 0.05 | 1.12 | 0.62 |
| L2 | snap23 | Synaptosomal-associated protein 23 | O09044 | 1.32 | 0.11 | 0.81 | 0.11 | 1.34 | 0.04 | 1.09 | 0.61 |
| **L3** | **snx6 ac.** | **Sorting nexin 6** | **Q6P8X1** | **0.74** | **0.10** | **0.47** | **0.02** | **0.66** | **0.02** | **0.63** | **0.06** |
| **L4** | **ssrd** | **Translocon-associated protein subunit delta** | **062186** | **1.15** | **0.15** | **1.24** | **0.04** | **1.07** | **0.49** | **1.27** | **0.04** |
| L5 | vps25 | Vacuolar protein-sorting-associated protein 25 | Q9CQ80 | 0.30 | 0.03 | 0.35 | 0.04 | 1.39 | 0.10 | 1.68 | 0.06 |
| L6a | anxa4 ac | Annexin A4 | P97429 | 1.18 | 0.02 | 1.02 | 0.71 | 1.76 | 0.01 | 1.45 | 0.01 |
| L6b | anxa4 bas | Annexin A4 | P97429 | 1.33 | 0.01 | 1.24 | 0.07 | 1.01 | 0.97 | 1.08 | 0.63 |
| L7 | snf8 | Vacuolar-sorting protein SNF8 | Q9CZ28 | 1.57 | 0.01 | 1.24 | 0.08 | 1.05 | 0.66 | 0.71 | 0.16 |
| L8 | nsf1c | NSFL1 cofactor p47 | Q9C244 | 0.82 | 0.10 | 0.61 | 0.01 | 1.43 | 0.01 | 1.04 | 0.85 |
| **Signalling (s)** | | | | | | | | | | | |
| s1 | cab39 ac | Calcium-binding protein 39 | Q06138 | 0.85 | 0.24 | 0.45 | 0.01 | 1.01 | 0.95 | 1.15 | 0.42 |
| s2 | gbb1 | Guanine nucleotide-binding protein G(I)/G(S)/G (T) subunit beta-1 | P62874 | 1.06 | 0.24 | 1.39 | 0.01 | 1.13 | 0.45 | 1.09 | 0.41 |
| s3 | ppp1ca | Serine/threonine-protein phosphatase PP1-alpha catalytic subunit | P62137 | 0.90 | 0.18 | 0.76 | 0.01 | 0.99 | 0.93 | 1.04 | 0.79 |
| s4 | 14-3-3 eps | 14-3-3 protein epsilon | P62259 | 1.05 | 0.57 | 0.83 | 0.01 | 1.59 | 0.02 | 1.17 | 0.42 |
| **s5** | **mk14** | **Mitogen-activated protein kinase 14** | **P47811** | **0.99** | **0.91** | **1.35** | **0.02** | **1.07** | **0.67** | **1.75** | **0.01** |
| s6 | arl3 | ADP-ribosylation factor-like protein 3 | Q9WUL7 | 1.23 | 0.31 | 0.51 | 0.02 | 1.18 | 0.35 | 0.94 | 0.79 |
| s7 | gbb2 | Guanine nucleotide-binding protein G(I)/G(S)/G (T) subunit beta-2 | P62880 | 0.98 | 0.76 | 1.17 | 0.04 | 1.23 | 0.02 | 1.13 | 0.25 |
| **s8** | **pde6d** | **Retinal rod rhodopsin-sensitive cGMP 3',5'-cyclic phosphodiesterase subunit delta** | **O55057** | **1.52** | **0.25** | **2.07** | **0.04** | **1.95** | **0.04** | **1.42** | **0.02** |
| s9 | cpped | Serine/threonine-protein phosphatase CPPED1 | Q8BFS6 | 0.63 | 0.06 | 0.60 | 0.04 | 1.07 | 0.65 | 0.92 | 0.36 |
| s10a | ptpn6 ac | Tyrosine-protein phosphatase non-receptor type 6 | P29351 | 0.72 | 0.14 | 0.59 | 0.05 | N.D. | N.D. | N.D. | N.D. |
| s10b* | ptpn6 bas | Tyrosine-protein phosphatase non-receptor type 6 | P29351 | 0.62 | 0.10 | 0.56 | 0.07 | N.D. | N.D. | N.D. | N.D. |
| *s11* | *ctbp1* | *C-terminal-binding protein 1* | *O88712* | *0.83* | *0.05* | *0.86* | *0.09* | *1.3* | *0.11* | *1.21* | *0.35* |
| **Cell death control (a)** | | | | | | | | | | | |
| a1a | bid ac. | BH3-interacting domain death agonist | P70444 | 0.64 | 0.17 | 0.31 | 0.01 | 1.49 | 0.03 | 1.39 | 0.04 |
| *a1b* | *bid bas.* | *BH3-interacting domain death agonist* | *P70444* | *1.32* | *0.01* | *1.06* | *0.63* | *1.13* | *0.48* | *1.23* | *0.37* |
| a2b | Casp 3 bas | Caspase-3 | P70677 | 0.71 | 0.09 | 0.26 | 0.01 | 0.74 | 0.52 | 1.90 | 0.06 |
| *a2a* | *casp3 mod* | *Caspase-3* | *P70677* | *0.77* | *0.02* | *1.01* | *0.92* | *0.87* | *0.18* | *0.89* | *0.53* |
| **Inflammation (i)** | | | | | | | | | | | |
| i1 | in35 | Interferon-induced 35 kDa protein homolog | Q9D8C4 | 0.80 | 0.03 | 0.59 | 0.01 | 1.15 | 0.09 | 1.08 | 0.64 |
| i2 | anxa1 | Annexin A1 | P10107 | 1.02 | 0.73 | 1.13 | 0.01 | 1.4 | 0.27 | 0.94 | 0.82 |
| i3 | myd88 | Myeloid differentiation primary response protein MyD88 | P22366 | 0.74 | 0.24 | 0.43 | 0.02 | 1.22 | 0.15 | 1.33 | 0.03 |

**Table 1** (Contd.)

| Spot Id. | Protein Short name | Protein Full name | Prot. acc. no. (uniprot) | Ratio RAW264.7 TMA10/ctl | t test RAW264.7 TMA10 vs. ctl | Ratio RAW264.7 TMA20/ctl | t test RAW264.7 TMA20 vs. ctl | Ratio MPC11 TMA20/ctl | t test MPC11 TMA20 vs. ctl | Ratio MPC11 TMA100/ctl | t test MPC11 TMA100/ TMA100 vs. ctl |
|---|---|---|---|---|---|---|---|---|---|---|---|
| i4 | **casp1** | **Caspase-1** | P29452 | 0.85 | 0.43 | 0.45 | 0.02 | 0.61 | 0.01 | 0.56 | 0.04 |
| Miscellaneous (u) | | | | | | | | | | | |
| u1 | **mtna** | **Methylthioribose-1-phosphate isomerase** | Q9CQT1 | 1.08 | 0.20 | 1.30 | 0.01 | 1.32 | 0.04 | 1.40 | 0.02 |
| u2 | cry1l | Lambda-crystalline homolog | Q99KP3 | 1.00 | 0.99 | 0.49 | 0.01 | 1.14 | 0.51 | 1.24 | 0.33 |
| u3 | ppac | Low molecular weight phosphotyrosine protein phosphatase | Q9D358 | 0.40 | 0.01 | 0.32 | 0.01 | 1.28 | 0.15 | 1.61 | 0.03 |
| u4a | **clic4 ac** | **Chloride intracellular channel protein 4** | Q9QYB1 | 1.26 | 0.05 | 1.19 | 0.06 | 1.13 | 0.17 | 1.25 | 0.01 |
| u4b | **clic4 bas** | **Chloride intracellular channel protein 4** | Q9QYB1 | 1.15 | 0.03 | 1.22 | 0.01 | 1.44 | 0.01 | 1.34 | 0.01 |
| u5 | lhpp | Phospholysine phosphohistidine inorganic pyrophosphate phosphatase | Q9D7I5 | 1.20 | 0.16 | 1.63 | 0.01 | 0.57 | 0.11 | 0.4 | 0.11 |
| u6 | amrp | Alpha-2-macroglobulin receptor-associated protein | P55302 | 0.98 | 0.92 | 1.64 | 0.01 | 0.78 | 0.29 | 0.75 | 0.35 |
| u7 | mat2b | Methionine adenosyltransferase 2 subunit beta | Q99LB6 | 0.74 | 0.11 | 0.79 | 0.03 | 0.94 | 0.73 | 0.62 | 0.17 |
| u9 | fa49b | Protein FAM49B | Q921 M7 | 0.85 | 0.47 | 0.48 | 0.04 | 1.80 | 0.01 | 1.39 | 0.17 |
| u10 | cdc42 | Cell division control protein 42 | P60766 | 0.67 | 0.14 | 0.48 | 0.04 | 1.16 | 0.07 | 1.07 | 0.54 |
| u11 | spre | Sepiapterin reductase | Q64105 | 0.99 | 0.95 | 0.74 | 0.04 | 1.14 | 0.23 | 1.08 | 0.62 |

Underlined: proteins that change in opposite directions in RAW264 and MPC11 cell lines. Bold: proteins that change significantly in both RAW264 and MPC11 cell lines. Italics: proteins that change significantly in RAW264 at the low dose but not at the high dose. *Proteins that do not change significantly but are included as controls.

RAW264 cell line at the $LD_{20}$. It can also be noted that most proteins are present in both the MPC11 and RAW264 cell lines, although their amounts may be different in the two lines. This feature has been observed in wider proteomic screens[74] and show that "housekeeping proteins" represent in fact the vast majority of proteins, at least those detected in proteomic screens.

### 3.3. Validation studies

The inclusion of small but reproducible protein changes means in turn that these changes cannot be validated easily by classical biochemical techniques at the protein expression level. For example, protein blotting often shows a technical variability well above 20%, and a response curve often lower than that of 2D electrophoresis, making this technique unsuitable for the validation of small fold changes. This renders functional validation even more necessary, to confirm the biological relevance of the proteomics-detected protein modulations.

**3.3.1. Enzyme activities.** In 2D gel-based proteomics, proteins are often represented by several spots, and it happens frequently that one spot is changed under the biological conditions of interest while the others are more or less constant. As the different spots correspond to modified forms of the protein and as protein modifications are known to modulate enzyme activities (*e.g.* acetylation[75]), the correspondence between spot variations and enzyme activities is far from obvious and must be verified. We carried out this verification on three enzymes, namely isocitrate dehydrogenase (spots e1a and e1b), lactoylglutathione lyase (spot h9) and biliverdin reductase (spots h3a, h3b and h4). The results, displayed in Table 2, show that the activity correlates with the spot change observed on 2D gels for lactoylglutathione lyase. In the case of isocitrate dehydrogenase, the activity correlates with the change in the acidic spot corresponding to the enzyme, and neither with the basic spot nor with the sum of the two spots. This example demonstrates, if further needed, the interest of a proteomic analysis taking into account the protein species and not only the gene product level. In the case of the biliverdin reductase activity, the situation is more complex. The best correlation is found with a combination of the acidic spot of biliverdin reductase A and the spot of biliverdin reductase B. This is in line with the known role of phosphorylation in the activity of biliverdin reductase A.[76]

**3.3.2. Cytoskeleton and phagocytosis.** Numerous proteins associated with the actin cytoskeleton emerged from the proteomic screen (see category c in Table 1). This led us to study if the actin cytoskeleton was altered in macrophages upon treatment with silica nanoparticles, using labelled phalloidin and confocal microscopy. The results, displayed in Fig. 3, show that silica nanoparticles induce a decrease in the number of spikes observed at the surface of the macrophages. This effect was not due to the internalization of particles *per se*, as internalization of latex particles did not induce this effect (Fig. 3, panels D and E).

**Table 2** Enzyme activities measured in control cell extracts and in extracts prepared from cells treated for 24 hours with either 10 µg ml$^{-1}$ or 20 µg ml$^{-1}$ silica nanoparticles. The activities are expressed in units per mg protein, the unit being defined as 1 µmol of substrate converted per minute

µmol per min per mg prot

| | Control | Silica (10 µg ml$^{-1}$) | Silica (20 µg ml$^{-1}$) |
|---|---|---|---|
| **Lgul** | | | |
| A | 364.9 | 329.3 | 334.7 |
| B | 356. | 316.9 | 300.9 |
| C | 370.3 | 336.5 | 238.6 |
| D | 366.7 | 349.0 | 306.2 |
| Mean | 364.5 | 332.9 | 295.1 |
| Std deviation | 6.06 | 13.40 | 40.5 |
| Fold change | | 0.91 | 0.81 |
| t test vs. control | | 0.011 | 0.040 |
| | | | |
| **IDHC** | | | |
| A | 21.5 | 23.5 | 14.5 |
| B | 19.5 | 29.5 | 16.5 |
| C | 31.5 | 21.0 | 20.5 |
| D | 33.5 | 27.5 | 11.0 |
| Mean | 26.5 | 25.37 | 15.62 |
| Std deviation | 7.02 | 3.84 | 3.97 |
| Fold change | | 0.96 | 0.59 |
| t test vs. control | | 0.791 | 0.045 |
| | | | |
| **Biliverdin reductase** | | | |
| A | 1.0 | 0.5 | 0.77 |
| B | 0.82 | 0.68 | 0.68 |
| C | 0.73 | 0.68 | 0.73 |
| D | 0.86 | 0.77 | 0.82 |
| Mean | 0.85 | 0.66 | 0.75 |
| Std deviation | 0.11 | 0.11 | 0.06 |
| Fold change | | 0.77 | 0.88 |
| t test vs. control | | 0.053 | 0.177 |

As the actin cytoskeleton is also involved in phagocytosis, we also tested this macrophage function. The results, displayed in Fig. 4A and B, show a moderate decrease in the proportion of phagocytic cells and an almost unchanged phagocytic ability for the phagocytosis-positive cell, for the cells exposed to 20 µg ml$^{-1}$ silica nanoparticles.

### 3.3.3. Mitochondrial potential.

Numerous mitochondrial proteins were found in the proteomic study (see category m in Table 1), some of them implied in energy generation directly (e.g. the long-chain specific acyl-CoA dehydrogenase (acadl, spot m9)), the beta subunit of the succinate-CoA ligase [GDP-forming] (sucb2, spots m1a and m1b) or indirectly through control of the oxphos complexes (e.g. the mitochondrial elongation factor Tu (eftu, spot m5), or the ubiquinone biosynthesis protein coq9 (spot m2)). This prompted us to assess the mitochondrial transmembrane potential. The results, displayed in Fig. 4C and D, show no alteration either in the proportion of cells with a normal transmembrane potential or on the value of the potential. This result is in line with the fact that most mitochondrial proteins picked in the proteomic screen show an increase of abundance. This illustrates the fact that the cells increase the amount of some mitochondrial proteins to compensate for the effects of the silica treatment, and are thus able to maintain the mitochondrial potential at the sub-toxic concentrations used.

### 3.3.4. Effects on signaling pathways.

Several proteins associated with signaling were found modulated upon silica nanoparticle treatment of macrophages, according to proteomics. We focused on the AMP-activated protein kinase (AMPK pathway) and on the myeloid differentiation primary response protein 88 (myd88) pathway.

The activity of the AMPK pathway is controlled by the STK11/LKB1 kinase,[77] whose activity is controlled through the formation of a ternary LKB1-STRAD-CAB39/Mo25 complex.[78] In this model, a decrease in the calcium binding protein 39 (CAB39/Mo25) should result in a decrease of the LKB1 activity, resulting in turn in a decrease of the AMPK activity. Macrophages respond to silica nanoparticles by decreasing the amount of CAB39/Mo25 (spot s1) and thus putatively decreasing the activity of the AMPK pathway. It is also worth noting that macrophages also respond to silica nanoparticles by a strong decrease of adenosine kinase (adk, spot r9) whose product is AMP, i.e. another activator of AMPK.

We thus tested if a pharmacological inhibition of AMPK would alter cell survival after treatment with silica nanoparticles. The results, displayed in Fig. 5A, show that inhibition of the AMPK pathway dramatically increases cell survival upon treatment with silica.

The myd88 pathway is involved in the transduction of the signals produced by activation of most Toll like receptors (TLR).[79] Consequently, a decrease in myd88, as we observed in our proteomic screen (spot i3), should result in a lesser efficiency of the TLR pathways and thus to decreased responses when the TLR are stimulated. To test this hypothesis, we used the classical lipopolysaccharide (LPS)-induced NO production, linked to the stimulation of TLR4. The results of these experiments, displayed in Fig. 5B, show a progressive decline of the LPS-induced NO production when cells are treated with silica nanoparticles. This effect was also observed for the production of interleukin 6, while Tumor Necrosis Factor alpha (TNF-alpha) did not show the same response (Fig. 5C and D).

### 3.3.5. Effects on DNA.

Among the proteins modulated in response to silica nanoparticles, a few are associated with DNA replication and DNA repair. We thus first checked if DNA damage was observable in silica treated RAW264.7 cells. The results, displayed in ESI Fig. 8,† show that treatment of RAW264.7 cells with 20 µg ml$^{-1}$ silica induces an increase in DNA damage, as previously detected in other cell types[6] and/or with larger silica particles.[24] One of the modulated proteins is the 2′-deoxynucleoside 5′-phosphate N-hydrolase 1 (DNPH1, spot d3), a protein cleaving off the base from deoxyribonucleotides, independently from the nature of the base, although the kinetics of the cleavage may differ from one nucleotide to another.[80–82] In the context of nucleotide excision repair, whose end products are nucleotides, the role of such an enzyme is to remove the damaged bases from the nucleotides, so that damaged nucleotides cannot re-enter the salvage pathway and be re-incorporated into DNA.

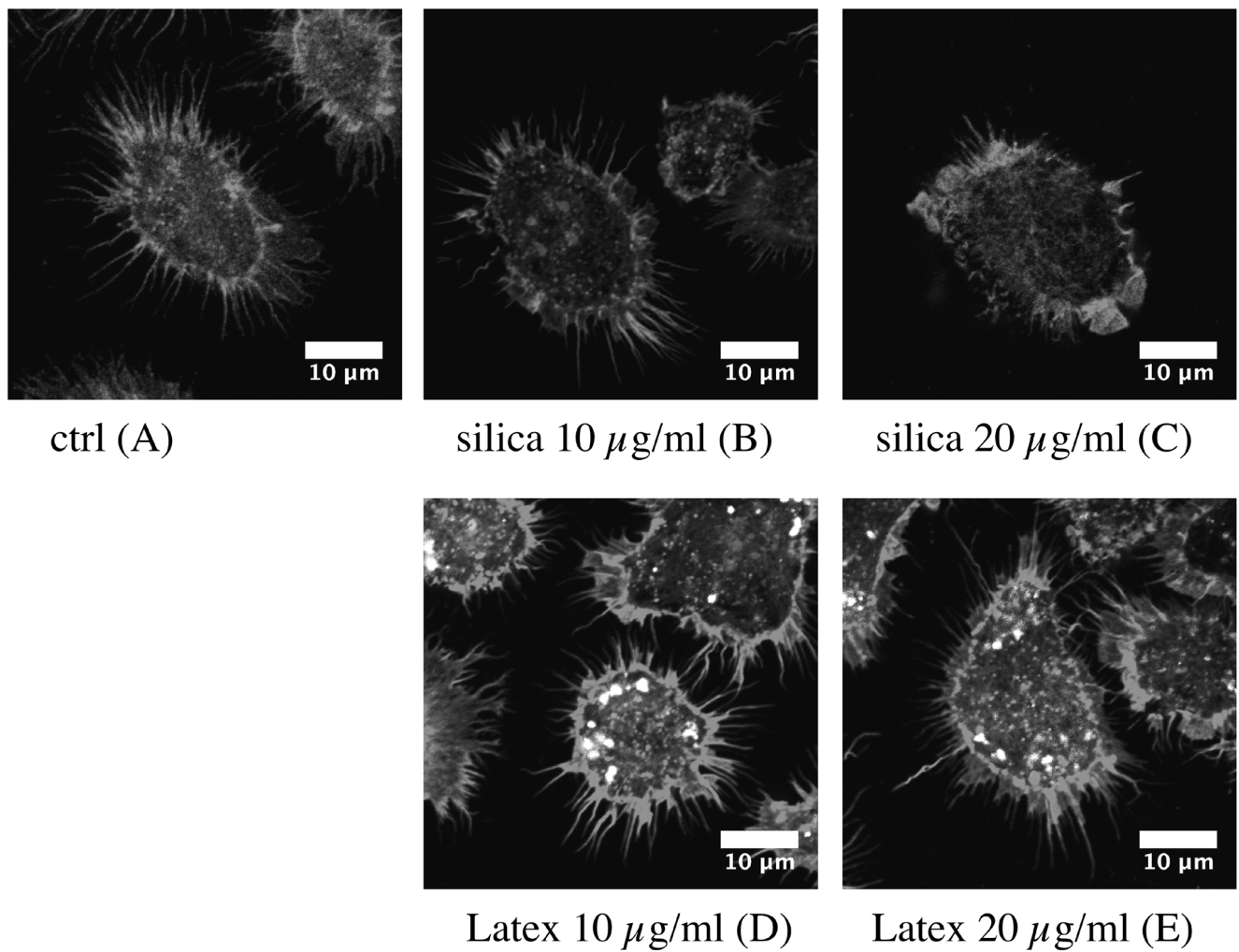

ctrl (A)    silica 10 $\mu$g/ml (B)    silica 20 $\mu$g/ml (C)

Latex 10 $\mu$g/ml (D)    Latex 20 $\mu$g/ml (E)

**Fig. 3** Changes in the cell morphology and actin cytoskeleton. Three dimensional reconstructions of the F-actin cytoskeleton (visualized by phalloidin staining) are shown, allowing visualization of the surface ruffles of the cells. Top row: Control cells and cells treated for 24 hours with Ludox TMA silica. Bottom row: Cells treated with fluorescent latex beads. Note the loss of surface spikes induced by silica but not by latex, showing that the effect is not solely induced by the phagocytosis process *per se*.

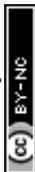

Another enzyme that can be implicated, although indirectly, in the control of DNA damage is lactoylglutathione lyase,[83] as it prevents glycation of nucleotides by destroying methylglyoxal. As guanine glycation has been already observed with zinc oxide nanoparticles,[56] we first checked if the observed decrease in the amount and activity of lactoylglutathione lyase (spot h9) was correlated with an increased sensitivity to methylglyoxal. The results, displayed in Fig. 6A and B, show no significant difference in methylglyoxal sensitivity between control cells and silica-treated cells.

If the hypothesis regarding the role of DNPH1 is correct, a decrease in this protein may mean a decrease in detoxification of damaged DNA and thus an increased sensitivity to DNA-damaging agents detoxified *via* the nucleotide excision repair (NER) pathway. To test this hypothesis, we examined the effect of silica nanoparticles on the cellular sensitivity to styrene oxide, a bulky nucleophilic agent inducing cell death and DNA damage in other models.[84,85] The results, displayed in Fig. 6C and D, show that pre-treatment of macrophages with silica nanoparticles induces an increased sensitivity to styrene oxide. This cross-toxicity effect is not present on MPC11 cells, which show a silica-induced increase in the cellular amount of DNPH1 instead of the decrease observed in macrophages. To obtain further insights into the alterations of the NER system induced by silica nanoparticles, we investigated the changes in the expression of some of the proteins of the system by RT-qPCR. The results, displayed in Fig. 7, show a decrease in the expression of some of these genes: Cockayne Syndrome Protein A (CSA), Proliferating Cell Nuclear Antigen (PCNA), DNA repair protein XRCC1, and DNA excision repair protein (ERCC1) upon macrophage treatment with silica nanoparticles at the $LD_{20}$ dose. Only two genes (CSA and XRCC) show a modulated response in the MPC11 cell line and here again at the $LD_{20}$ dose.

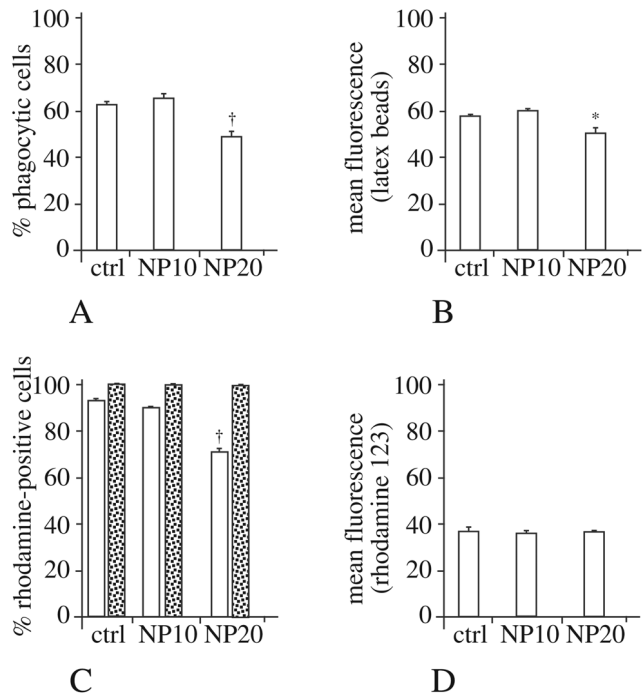

**Fig. 4** Study of the phagocytic index and of the mitochondrial transmembrane potential. Panel A: Proportion of phagocytic cells (in the viable cell population only) for control cells or cells treated for 24 hours with 10 or 20 µg ml⁻¹ Ludox TMA silica. Panel B: Mean fluorescence of phagocytic cells (in the viable cell population only) for control cells or cells treated for 24 hours with 10 or 20 µg ml⁻¹ Ludox TMA silica. Panel C: Proportion of Rhodamine 123-positive cells in the total population (white bars) or in the viable cell population only (dotted bars) for control cells or cells treated for 24 hours with 10 or 20 µg ml⁻¹ Ludox TMA silica. Panel D: Mean Rhodamine 123 fluorescence (in the viable cell population only) for control cells or cells treated for 24 hours with 10 or 20 µg ml⁻¹ Ludox TMA silica. Symbols indicate the statistical significance (Student's t-test): *$p < 0.05$; †$p < 0.01$.

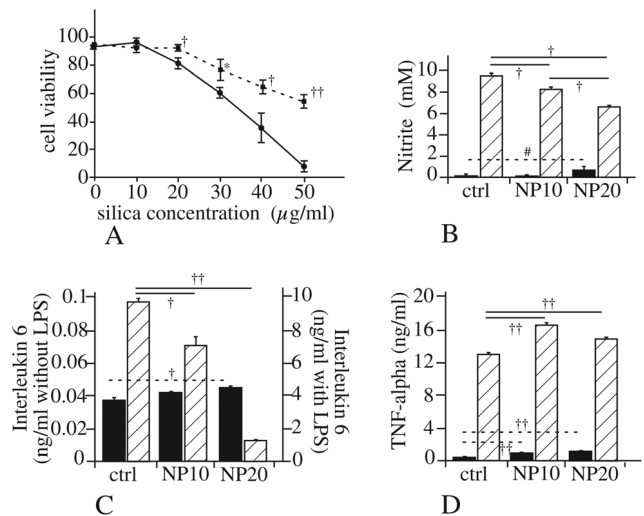

**Fig. 5** Validation of the effects on signaling pathways. Panel A: Dose dependent survival curve for RAW264.7 cells treated for 24 hours with various concentrations of Ludox TMA silica, with (dotted lines) or without co-treatment with the PKA inhibitor H89 (10 µM). Panel B: NO production of RAW264.7 cells treated with Ludox TMA nanoparticles only (black bars) or treated with nanoparticles and 100 ng ml⁻¹ lipopolysaccharide (hatched bars). #Significant difference at $p < 0.05$ with the Mann–Whitney U test but not the Student's t-test ($p = 0.08$). †$p < 0.01$ according to the Student's t-test. Panel C: IL6 production of RAW264.7 cells treated with Ludox TMA nanoparticles only (black bars) or treated with nanoparticles and 100 ng ml⁻¹ lipopolysaccharide (hatched bars). Symbols indicate the statistical significance (Student's t-test): *$p < 0.05$; †$p < 0.01$; ††$p < 0.001$. Note the different scales for IL6 production for cells treated with silica alone (left scale) or with silica and LPS (right scale). Panel D: TNF-alpha production of RAW264.7 cells treated with Ludox TMA nanoparticles only (black bars) or treated with nanoparticles and 100 ng ml⁻¹ lipopolysaccharide (hatched bars). Symbols indicate the statistical significance (Student's t-test): *$p < 0.05$; †$p < 0.01$; ††$p < 0.001$.

## 4. Discussion

One of the major problems in the field of nanotoxicology is the variability of the results presented in the scientific literature. The variability of the nanomaterials, even if they bear the same chemical name, and their poor characterization have often been blamed as the main cause of this observed variability. However, in many cases both the cell types and the nanoparticles used change from one study to another, and it is difficult to evaluate the influence of each factor on the final variability. Even worse, some cell lines such as those used in the present study can be cultured in two different media, namely RPMI 1640 and DMEM. When comparing our results with those obtained by Panas et al.[7] on the same cell line with precipitated amorphous silica of very similar size (26 vs. 25 nm), we found a much higher cytotoxicity than they did. However our cells are grown in RPMI 1640 while in their study the cells were grown in DMEM, and it has been recently demonstrated that such a medium change induces significant changes in the cellular proteome and in the observed responses to nanoparticles.[86] This may be linked to the already-described dependence of silica toxicity on metabolic activity,[87] which may be different between the rich DMEM medium and the relatively poorer RPMI 1640 medium.

This being stated, the increased sensitivity of macrophages to amorphous silica, which has been established in several studies, both for nanoparticles e.g. ref. 5–7 and microparticles[12] is further confirmed in the present study.

In the case of amorphous silica, one factor that greatly affects cytotoxicity is the presence of a protein corona, which forms when silica is introduced into a protein-containing medium such as culture media with bovine serum.[88] It has been shown that the presence of the corona decreases the toxicity of silica (e.g. in ref. 89 and 90). This means that in the presence of proteins, the entity that is internalized is not a bare silica particle, but a core–shell silica–protein particle, which means in turn that in this case, adsorbed proteins are introduced into the cell. It may be then questioned whether such internalized proteins may affect cellular physiology. Although such a hypothesis cannot strictly be ruled out, it seems unlikely for two reasons. First the proteins adsorbed on nanoparticles are often denatured,[91,92] which means that they lose their functions. Second, silica nanoparticles are internalized in lysosomes[14] which are in an acidic environment where massive protein degradation occurs. Consequently, a functional effect of the corona inside the cells seems rather unlikely.

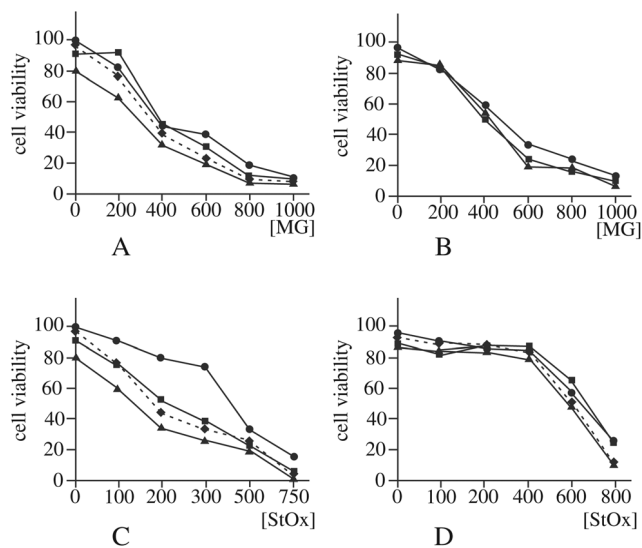

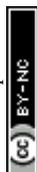

**Fig. 6** Validation of cross toxic effects. In these experiments the cells were pre-treated with 0, 10 or 20 μg ml$^{-1}$ Ludox TMA nanoparticles for 6 hours, then the toxic agent was added for a further 18 hours. Cell viability was assessed by dye exclusion after this total 24 hour treatment. Panel A: Survival curve for RAW264.7 cells co treated with silica and methylglyoxal (MG). Circles: Control cells (not treated with silica). Squares: Cells treated with 10 μg ml$^{-1}$ silica. Triangles: Cells treated with 20 μg ml$^{-1}$ silica. Diamonds and dotted line: Cells treated with 20 μg ml$^{-1}$ silica, with correction of the mortality induced by silica alone. No statistically significant effect can be detected for cells treated with 10 μg ml$^{-1}$ silica, and after correction of the mortality induced by silica alone, a moderate but statistically significant effect ($p < 0.05$) can be detected for cells treated with 20 μg ml$^{-1}$ silica only at 600 μM methylglyoxal. Panel B: Same as panel A, but for MPC11 cells. No statistically significant effect could be detected. Panel C: Survival curve for RAW264.7 cells co-treated with silica and styrene oxide (StOx). Circles: Control cells (not treated with silica). Squares: Cells treated with 10 μg ml$^{-1}$ silica. Triangles: Cells treated with 20 μg ml$^{-1}$ silica. Diamonds and dotted line: Cells treated with 20 μg ml$^{-1}$ silica, with correction of the mortality induced by silica alone. Except for the 0 μM styrene oxide point, all points are statistically different between the cells treated with 20 μg ml$^{-1}$ silica ($p < 0.01$ for 100, 200, 300 μM styrene oxide, $p < 0.05$ for 500 and 750 μM styrene oxide) and cells treated with styrene oxide alone, even after correction of the mortality induced by silica alone. For the cells treated with 10 μg ml$^{-1}$ silica, all points except the 0 μM styrene oxide point are statistically different from the control ($p < 0.01$). Panel D: Same as panel C, but for MPC11 cells. No statistically significant effect could be detected.

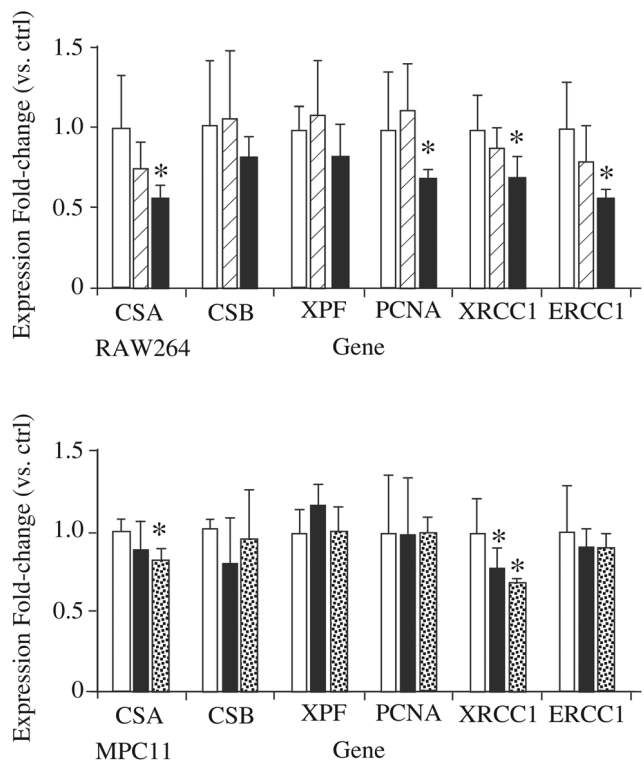

**Fig. 7** Expression analysis by RT-qPCR of genes involved in the nucleotide excision repair (NER) pathway. The expression of six genes involved in the NER pathway was monitored by RT-qPCR for RAW264.7 and MPC11 cells treated with Ludox TMA silica nanoparticles. White bars: Control cells. Hatched bars: Cells treated for 24 hours with 10 μg ml$^{-1}$ silica. Black bars: Cells treated for 24 hours with 20 μg ml$^{-1}$ silica. Dotted bars: Cells treated for 24 hours with 100 μg ml$^{-1}$ silica. *$p < 0.05$ in the Student's $t$-test.

As another example of the variability of the results published in the literature, proteomic studies on the cellular responses to silica nanoparticles have been published previously[42,93] and have shown widely different results and even results different from those presented in the present study. However, the low number of common responses detected in the present study between two different cell lines at equal effect dose or at identical doses clearly demonstrates that cellular responses are widely different from one cell type to another, and such cell type-specific effects are the main explanation for the different responses reported using proteomic studies.

There are however discrepancies that remain, even when working with the same cell line in the same culture medium and with similar nanoparticles. One example can be found in the synergy between silica and LPS for NO production. In our study we found a negative synergy (silica decreases the LPS-induced NO production) while Di Cristo *et al.* found a positive synergy.[30] Several factors can explain this discordance in the results: (i) the duration of the LPS treatment (24 hours in our case instead of 48 hours in the Di Cristo *et al.* study), (ii) the concentration of silica used (10 μg cm$^{-2}$ in the Di Cristo *et al.* study instead of 4 μg cm$^{-2}$ in our case) and (iii) the fact that in this precise case Di Cristo *et al.* used a suboptimal LPS concentration (10 ng ml$^{-1}$) instead of the 100 ng ml$^{-1}$ concentration that we used and also gave full NO production in the Di Cristo *et al.* study. While the duration of the treatment may explain the differences in the absolute NO concentration found (20 mM in their case instead of 9.5 mM in our study), the opposite synergies found probably depend on the last two factors. Indeed, a positive synergy will be impossible to observe if the maximal stimulation is already reached. In addition, the silica dose used in the Di Cristo *et al.* study is fairly important and induces strong membrane permeability (50% at 24 hours). As this parameter is the one tested by the viability assay that we used (trypan blue exclusion), such a condition corresponds to the $LD_{50}$ in our system. Such an activation of the proinflammatory functions of macrophages at toxic concentrations has already been described for silica[11] and for other nanoparticles,[94] and may contribute to explain the differences between the two sets of results on the silica-LPS synergy.

In order to extend our observations on the TLR axis, we also measured the release of the two inflammatory cytokines IL6 and TNF. While IL6 showed the same response as NO production (weak induction by silica alone and strong reduction by silica of the LPS-induced response) TNF did not show the same response pattern. This discordance between the two cytokine responses has been observed previously in the case of copper oxide nanoparticles.[48]

One of the more interesting outcomes of the proteomic analysis lies in the proteins involved, directly or indirectly, in the DNA repair pathway. Such proteins include PCNA, DNPH1 and lactoylglutathione lyase. Regarding lactoylglutathione lyase, we did not observe any cross toxicity between silica and methylglyoxal, opposite to what was observed with zinc nanoparticles.[56] The decrease of PCNA and DNPH1 suggested however a decrease in the efficiency of the NER pathway, which resulted in a higher sensitivity to bulky DNA alkylating agents. This sensitivity was however observed only on the sensitive cell type (macrophage) and not on the less sensitive MPC11 cell line. Such a cross sensitivity may be relevant for pulmonary toxicity, as it could induce a decrease in the number of viable lung macrophages if they are exposed both to silica nanoparticles and to DNA alkylating agents, such as those contained in tobacco smoke or combustion particles at a larger sense. A similar cross toxicity between insoluble nanoparticles and DNA alkylating agents has been previously observed with titanium dioxide.[95] Genotoxicity has been previously described for silica nanoparticles,[25,27] but has been observed only at high, cytotoxic concentrations.

These cross toxic effects between nanoparticles and chemicals have been described with metallic ions such as cadmium[39] and lead,[40] but our work and the one of Armand *et al.* extend it to organic chemicals. The sequential treatment used (nanoparticles first, then chemicals) is not in favor of a direct trojan horse effect, *i.e.* an adsorption of the chemical on the nanoparticles leading to a better penetration in the cells and an intracellular release of the adsorbed chemical. It is more in favor of a synergistic effect, *i.e.* an alteration of the cellular physiology by the nanoparticle which renders the cell more sensitive to the chemical of interest.

Such studies of cross effects are important in a safe by design perspective. Primary determinants of cellular toxicity are of course of crucial importance, and have been recently described for silica.[20] They are however not sufficient, as the real use of the products involves co-exposures that are difficult to predict and may vary greatly, *e.g.* according to lifestyle. In such a frame, wide-scope studies such as omics studies are able to provide valuable insights, provided that they are fully interpreted down to the protein level and not only to the pathway level, and provided that their predictions are tested.

## Author contributions

BD and CAG performed the DLS, phagocytosis, NO and the mitochondrial potential experiments. In addition BD performed the F-actin staining. MC, FD and DB performed the RT-qPCR and the comet experiments. HD and SC performed and interpreted the mass spectrometry identification in the proteomics experiments, and helped in drafting the manuscript.

DF and GS performed and interpreted the TEM experiments on the nanomaterial.

VCF and TR performed the 2D gel electrophoresis and enzyme assay experiments. TR performed the cross-toxicity experiments. In addition TR conceived and designed the whole study and drafted the manuscript. MC and CAG helped in designing the whole study and in drafting the manuscript, and critically revised the manuscript. All authors critically read and approved the manuscript.

## Abbreviations

| AMP | Adenosine 5′ monophosphate |
|---|---|
| CHAPS | 3-[(3-Cholamidopropyl)dimethylammonio]-1-propanesulfonate |
| DIGE | Differential in-gel electrophoresis |
| DMEM | Dulbecco modified Eagle's medium |
| EDTA | Ethylene diamine $N,N,N',N'$-tetraacetic acid |
| EGTA | Ethylene glycol-bis(2-aminoethylether)-$N,N,N',N'$-tetraacetic acid |
| FBS | Fetal bovine serum |
| $LD_{20}$ | Lethal dose 20% |
| LPS | Lipopolysaccharide |
| NADPH | Nicotinamide adenine dinucleotide phosphate, reduced form |
| PBS | Phosphate buffered saline |
| RPMI | Rockwell Park Memorial Institute |
| RT-qPCR | Reverse transcriptase-quantitative polymerase chain reaction |
| SDS | Sodium dodecyl sulfate |
| TCEP | Tris(carboxyethyl) phosphine |

## Acknowledgements

This work was funded by the CNRS, The University of Grenoble, the University of Strasbourg Unistra, the Région

Alsace, the French National Research Program for Environmental and Occupational Health of ANSES (PNREST 2015/032, Silimmun Grant), the toxicology project of the CEA (Imaginatox grant), and the French National Research Agency (ANR-16-CE34-0011, Paipito grant).

This work is a contribution to the Labex Serenade (no. ANR-11-LABX-0064) funded by the "Investissements d'Avenir" French Government program of the French National Research Agency (ANR) through the A*MIDEX project (no. ANR-11-IDEX-0001-02).

This work used the platforms of the Grenoble Instruct centre (ISBG; UMS 3518 CNRS-CEA-UJF-EMBL) with support from FRISBI (ANR-10-INSB-05-02) and GRAL (ANR-10-LABX-49-01) within the Grenoble Partnership for Structural Biology (PSB), as well as the platforms of the French Proteomic Infrastructure (ProFI) project (grant ANR-10-INBS-08-03). The electron microscope facility is supported by the Rhône-Alpes Region, the Fondation Recherche Medicale (FRM), the fonds FEDER, the Centre National de la Recherche Scientifique (CNRS), the CEA, the University of Grenoble, EMBL, and the GIS-Infrastrutures en Biologie Sante et Agronomie (IBISA). We also thank the Fondation pour la Recherche Médicale for financial support of a Synapt HDMS mass spectrometer.

Last but certainly not least, BD thanks the CNRS for a handicap PhD fellowship.